\documentclass[twocolumn,showkeys,aps,prb,showpacs]{revtex4-1}
\usepackage{graphicx}
\usepackage[CJKbookmarks,dvipdfm,colorlinks,linkcolor=blue,citecolor=blue]{hyperref}

\begin{document}

\title{Biaxial strain tuned thermoelectric properties  in  monolayer $\mathrm{PtSe_2}$}

\author{San-Dong Guo  and Lun Zhang}
\affiliation{Department of Physics, School of Sciences, China University of Mining and
Technology, Xuzhou 221116, Jiangsu, China}
\begin{abstract}
Strain engineering is a very effective method to tune electronic, optical, topological and thermoelectric properties of materials. In this work, we systematically study biaxial strain dependence of electronic structures and  thermoelectric properties (both electron and phonon parts) of   monolayer $\mathrm{PtSe_2}$ with generalized gradient approximation  (GGA) plus spin-orbit coupling (SOC) for electron part and GGA for phonon part. Calculated results show that compressive or tensile strain can induce conduction band minimum (CBM) or valence band maximum (VBM) transition, which produces important effects on Seebeck coefficient. It is found that compressive or tensile strain can induce  significantly enhanced n- or p-type Seebeck coefficient   at the critical strain of CBM or VBM transition, which  can be explained by  strain-induced  band convergence. Another essential strain effect is that  tensile strain can produce significantly reduced lattice thermal conductivity, and the room temperature lattice thermal conductivity at the strain of -4.02\% can decrease by about 60\% compared to unstrained one, which is very favorable for high $ZT$. To estimate efficiency of thermoelectric conversion, the  figure of merit $ZT$ can be obtained by empirical scattering time $\tau$. Calculated $ZT$ values show that
strain indeed is a very  effective strategy to achieve enhanced thermoelectric properties, especially for p-type doping.
Tuning thermoelectric properties with strain also can be applied to other semiconducting transition-metal dichalcogenide monolayers $\mathrm{MX_2}$ (M=Zr, Hf, Mo, W and Pt; X=S, Se and Te).

\end{abstract}

\pacs{72.15.Jf, 71.20.-b, 71.70.Ej, 79.10.-n ~~~~~~~~~~~~~~~~~~~~~~~~~~~~~~~~~~~Email:guosd@cumt.edu.cn}

\maketitle

\section{Introduction}
Thermoelectric materials have enormous potential to solve energy issues, since they  can realize direct hot-electricity conversion without moving parts by using the Seebeck effect and Peltier effect.
As is well known, the dimensionless  figure of merit\cite{s1,s2}, $ZT=S^2\sigma T/(\kappa_e+\kappa_L)$, can measure the efficiency of thermoelectric conversion, in which  S is the Seebeck coefficient,   $\sigma$ is electrical conductivity,  T is absolute  temperature,  $\kappa_e$ and $\kappa_L$ are the electronic and lattice thermal conductivities, respectively.
Bismuth-tellurium systems,   lead chalcogenides   and silicon-germanium alloys are  the most efficient for practical  application of thermoelectric devices\cite{s3,s4,s5}. According to the expression of $ZT$, high power factor ($S^2\sigma$) and low thermal conductivity ($\kappa=\kappa_e+\kappa_L$) can give rise to excellent efficiency of thermoelectric conversion, but often it is  to enhance one, while  adversely to affect  another.
Many  recent advances in improving efficiency of thermoelectric conversion are focused on low-dimensional materials due to
simultaneously increasing  power factor and decreasing thermal conductivity\cite{dw},  such as $\mathrm{Bi_2Te_3}$ nanowire,
monolayer phosphorene and silicene\cite{s11,s12,s13}.

\begin{figure}
  \includegraphics[width=7.0cm]{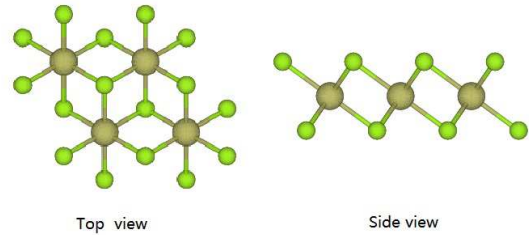}
  \caption{(Color online) The schematic crystal structure of monolayer $\mathrm{PtSe_2}$. The large  balls represent Pt atoms, and small balls Se.}\label{t0}
\end{figure}
\begin{figure*}
  \includegraphics[width=12cm]{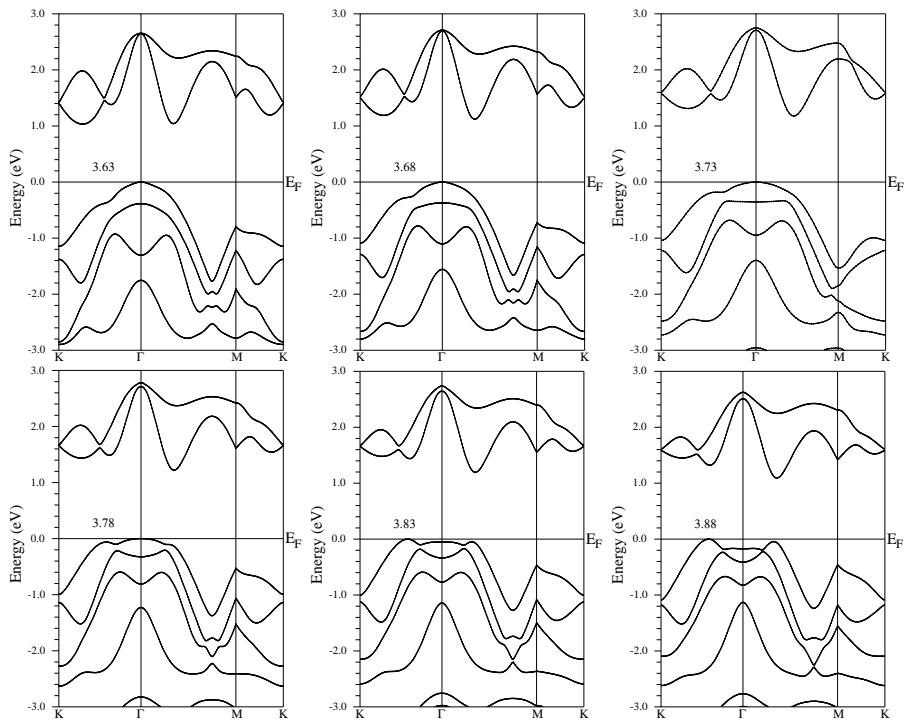}
\caption{The energy band structures  of monolayer $\mathrm{PtSe_2}$ with $a$ changing from 3.63 $\mathrm{{\AA}}$ to 3.88 $\mathrm{{\AA}}$  using GGA+SOC. }\label{t1}
\end{figure*}
\begin{figure}
  \includegraphics[width=7.0cm]{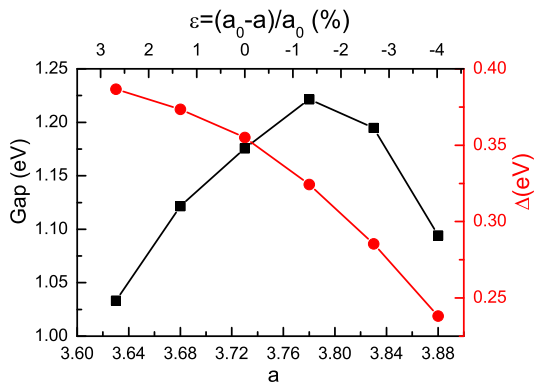}
  \caption{(Color online) The energy band gap (Gap) and  spin-orbit splitting value at $\Gamma$ point ($\Delta$) as a function of $a$ or $\varepsilon$ by using GGA+SOC.}\label{t2}
\end{figure}

 Due to the presence of  intrinsic  band gap, semiconducting two-dimensional (2D) transition-metal dichalcogenide  monolayers have more potential application in nanoelectronics and nanophotonics   in comparison with  the first 2D gapless Graphene.
The $\mathrm{MoS_2}$ of them is hot spot of present research both experimentally and theoretically \cite{q1,q2,q4,q6}, which has been applied in field effect transistors, photovoltaics and photocatalysis\cite{q7,q10,q11}.
Recently, the thermoelectric properties of transition-metal dichalcogenide  monolayers have attracted much attention\cite{t1,t2,t3,t4,t5,t6,gsd1,gsd2}. Thermoelectric performance of $\mathrm{MX_2}$ (M=Mo, W; X=S, Se) monolayers have been investigated using ab-initio method and ballistic transport model\cite{t1}, and at room temperature, a maximum ZT of monolayer $\mathrm{MoS_2}$  is obtained as 0.5. Experimentally, a value of S as 30 mV/K  has been reported for monolayer $\mathrm{MoS_2}$\cite{t2}, which is favorable for potential thermoelectric applications. Thermoelectric response of monolayer $\mathrm{MoSe_2}$ and $\mathrm{WSe_2}$  also have been studied by first-principles calculations and semiclassical Boltzmann
transport theory\cite{t6}.

Recently, we investigated spin-orbit and strain effect on power factor in monolayer $\mathrm{MoS_2}$\cite{gsd1}, and further systematically studied SOC effect on power factor in semiconducting transition-metal dichalcogenide monolayers $\mathrm{MX_2}$ (M=Zr, Hf, Mo, W and Pt; X=S, Se and Te)\cite{gsd2}. Among all cation groups, $\mathrm{PtX_2}$  (X=S, Se and Te) show the highest Seebeck coefficient, leading to best power factor, which indicates an great potential to attain excellent thermoelectric applications. The  monolayer $\mathrm{PtSe_2}$ of them  has been  epitaxially grown by direct selenization of Pt with high-quality single-crystal, which has  potential applications in valleytronics\cite{e1}. Moreover, the local Rashba spin polarization and
spin-layer locking in centrosymmetric monolayer  $\mathrm{PtSe_2}$ have been observed  by using spin- and angle-resolved photoemission spectroscopy, which has potential  applications in electrically tunable spintronics\cite{e2}.
 The first-principles calculations show that the band gaps of monolayer  $\mathrm{PtSe_2}$ can be tuned over a wide range by strain engineering\cite{e3}, but SOC is neglected, which has important effects on electronic structures of monolayer  $\mathrm{PtSe_2}$.

Here, the biaxial strain dependence of electronic structures and  thermoelectric properties of monolayer  $\mathrm{PtSe_2}$ are studied. The electron part is calculated using GGA+SOC, and it is very crucial to include SOC for attaining reliable power factor\cite{gsd1,gsd2}. Calculated results show that the energy band gap first increases, and then decreases with increasing lattice constants, while the spin-orbit splitting at $\Gamma$ point monotonically decreases. Compressive strain can induce CBM transition, while tensile strain can lead to VBM transition. The n- or p-type Seebeck coefficient can be significantly improved
at the boundary of CBM or VBM transition, which can be understood  by strain-induced  accidental degeneracies. It is found that
tensile strain can induce reduced lattice thermal conductivity. Finally, the $ZT$ values are attained, which shows strain indeed
can achieve enhanced thermoelectric properties.

The rest of the paper is organized as follows. In the next section, we shall
describe computational details. In the third section, we shall present strain dependence of the electronic structures and  thermoelectric properties of  monolayer  $\mathrm{PtSe_2}$. Finally, we shall give our discussions and conclusion in the fourth
section.

\begin{figure}
    \includegraphics[width=7cm]{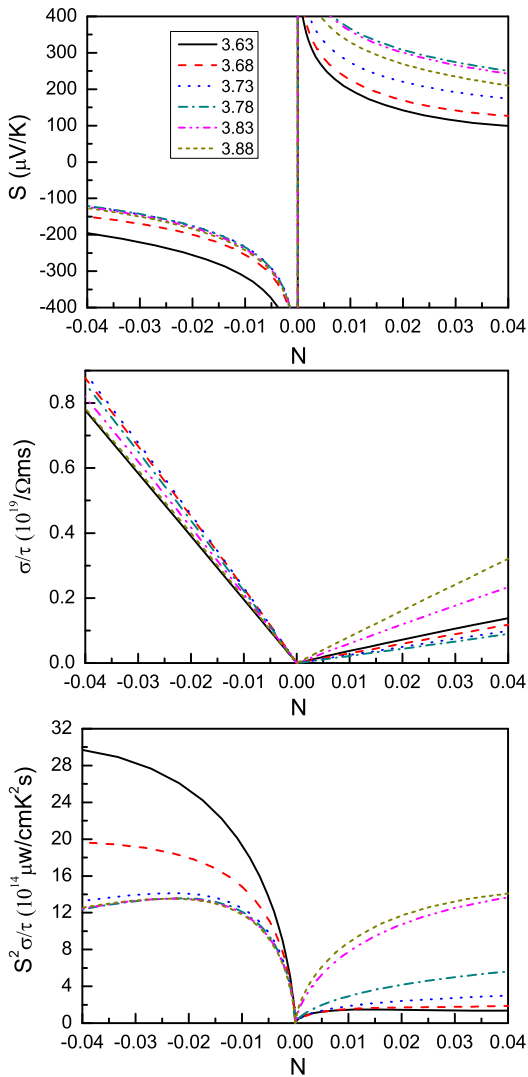}
  \caption{(Color online) At 300 K, transport coefficients, including  Seebeck coefficient S (Top), electrical conductivity with respect to scattering time  $\mathrm{\sigma/\tau}$ (Middle) and   power factor with respect to scattering time $\mathrm{S^2\sigma/\tau}$ (Bottom),  as a function of doping level (N) with $a$ changing from 3.63 $\mathrm{{\AA}}$ to 3.88 $\mathrm{{\AA}}$  using GGA+SOC.  The doping level (N) means  electrons (minus value) or holes (positive value) per unit cell.}\label{t3}
\end{figure}

\section{Computational detail}
The strain dependence of  electronic structures  of monolayer  $\mathrm{PtSe_2}$ is performed using
a full-potential linearized augmented-plane-waves method
within the density functional theory (DFT) \cite{1}, as implemented in
the WIEN2k  package\cite{2}.  We employ the popular GGA\cite{pbe} for the
exchange-correlation potential  to do our electron part calculations.
The  internal position parameters with a force standard of 2 mRy/a.u. are optimized using GGA.
 The SOC was included self-consistently \cite{10,11,12,so} due to containing heavy elements, which produces important influnces on power factor. To attain reliable results, we use 6000 k-points in the
first Brillouin zone for the self-consistent calculation,  make harmonic expansion up to $\mathrm{l_{max} =10}$ in each of the atomic spheres, and set $\mathrm{R_{mt}*k_{max} = 8}$. The self-consistent calculations are
considered to be converged when the integration of the absolute
charge-density difference between the input and output electron
density is less than $0.0001|e|$ per formula unit, where $e$ is
the electron charge. Transport calculations, such as  Seebeck coefficient, electrical conductivity and electronic
 thermal conductivity, are performed through solving Boltzmann
transport equations within the constant
scattering time approximation (CSTA) as implemented in
BoltzTrap\cite{b}, and reliable results  have
been obtained for several materials\cite{b1,b2,b3}. The accurate transport coefficients need dense k-point meshes, and we use 190 $\times$ 190 $\times$ 1 k-point meshes  in the
first Brillouin zone for the energy band calculation. The  lattice thermal conductivities are calculated within  the linearized phonon Boltzmann equation,  which can be achieved by using Phono3py+VASP codes\cite{pv1,pv2,pv3,pv4}. For the third-order force constants, 3$\times$3$\times$1 supercells
are built, and reciprocal
spaces of the supercells are sampled by  8$\times$8$\times$1 meshes. To compute lattice thermal conductivities, the
reciprocal spaces of the primitive cells  are sampled using the 20$\times$20$\times$1 meshes.

\section{MAIN CALCULATED RESULTS AND ANALYSIS}
The single-layer $\mathrm{PtSe_2}$ contains three atomic sublayers with Pt layer sandwiched between two Se layers, and the schematic crystal structure is shown in \autoref{t0}, which is different from crystal structure of  $\mathrm{MoS_2}$ due  to
different stacking of top and bottom Se or S sublayers.
The unit cell  of  monolayer $\mathrm{PtSe_2}$  contains  one Pt and two Se atoms, which is constructed with the
vacuum region of more than 15 $\mathrm{{\AA}}$ to avoid spurious interaction between neighboring layers, and the optimized lattice constant is $a$=3.73 $\mathrm{{\AA}}$  using GGA, which is very close to the
experimental value of 3.70 $\mathrm{{\AA}}$\cite{e1} or other theoretical value of 3.75  $\mathrm{{\AA}}$\cite{e3,e4}.
 The SOC has very important effects on electronic structures and thermoelectric properties, so SOC is included in all  calculations of electronic part except lattice part. The energy band structures with the optimized lattice constant $a$=3.73 $\mathrm{{\AA}}$ is plotted in \autoref{t1}, and calculated results show that $\mathrm{PtSe_2}$ is an indirect gap semiconductor with a band gap of 1.18 eV. The VBM is located at $\Gamma$ point,  while the CBM appears between $\Gamma$ and M points. The
first three valence bands near the $\Gamma$  point  are dominated by the Se-p character states, and the fourth valence band
is  mostly contributed by the Pt-d states. Due to  both inversion and time-reversal symmetries of $\mathrm{PtSe_2}$, all the bands are doubly degenerate.
\begin{figure}
  \includegraphics[width=7cm]{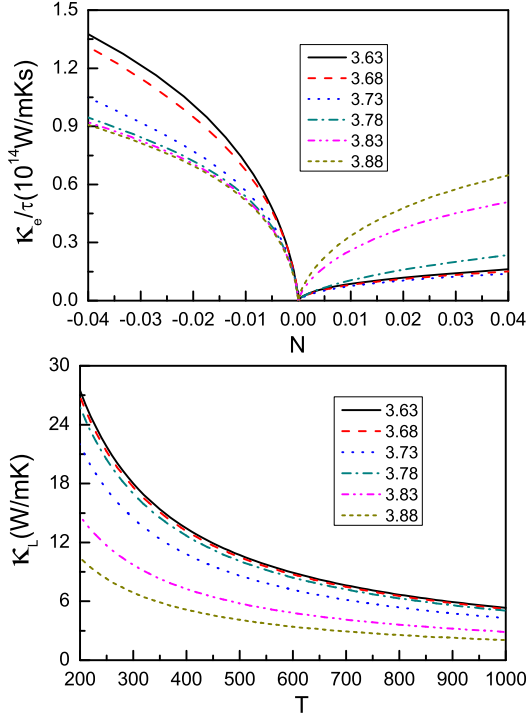}
  \caption{(Color online) The electronic thermal conductivity  with respect to scattering time  $\mathrm{\kappa_e/\tau}$  as a function of doping level  and lattice thermal conductivity  $\mathrm{\kappa_L}$  as a function of temperature with  $a$ changing from 3.63 $\mathrm{{\AA}}$ to 3.88 $\mathrm{{\AA}}$. }\label{t4}
\end{figure}

Both theoretically and experimentally, strain influence on the electronic structures and power factor  of monolayer $\mathrm{MoS_2}$ has been widely studied\cite{r45,r46,r47,gsd1}. Here, we investigate biaxial strain effects on the electronic structures and thermoelectric properties of  monolayer $\mathrm{PtSe_2}$.  The $\varepsilon=(a_0-a)/a_0$ is defined to simulate biaxial strain, and  $a_0$ is the optimized value of 3.73  $\mathrm{{\AA}}$ using GGA. $\varepsilon$$>$0 means  compressive strain, and $\varepsilon$$<$0 implies tensile strain. Biaxial strain dependence of  energy band gap  and   spin-orbit splitting value at $\Gamma$ point in the valence bands around the Fermi level using GGA+SOC  are plotted in \autoref{t2}, and the related energy band structures with six  considered $a$ values are also shown  in \autoref{t1}. As the $a$ increases, the energy band gap  firstly increases, and then decreases, which is similar to strain dependence of  monolayer $\mathrm{MoS_2}$\cite{gsd1}. The compressive strain leads to the  the transition of CBM from one point of $\Gamma$-M line  to one  point of K-$\Gamma$ line, while the VBM changes from $\Gamma$ point to one  point of K-$\Gamma$ line by applied tensile strain. The corresponding strain of CBM or VBM  transition is very small, which is about  $2.68\%$. These strain effects on electronic structures produce very important influences on power factor of  monolayer $\mathrm{PtSe_2}$.
As $a$ increases, the spin-orbit splitting at $\Gamma$ point monotonically decreases, and the change is  about 0.15 eV with $a$ varying  from 3.63 $\mathrm{{\AA}}$  to 3.88 $\mathrm{{\AA}}$. The spin-orbit splitting trend of monolayer $\mathrm{PtSe_2}$ is opposite to one of $\mathrm{MoS_2}$ with increasing $a$, and the spin-orbit splitting has stronger dependence on strain than one of $\mathrm{MoS_2}$\cite{gsd1}.

We  perform transport coefficients calculations, such as Seebeck coefficient S and electrical conductivity with respect to scattering time  $\mathrm{\sigma/\tau}$,  based on CSTA Boltzmann theory. The rigid band approach is employed, which is effective for low doping level\cite{tt9,tt10,tt11}. The doping effects on the  transport coefficients are simulated by changing the position of Fermi level.
As the Fermi level moves into conduction bands, the n-type doping is achieved with negative doping levels, giving the negative Seebeck coefficient.
The positive doping levels, giving the positive Seebeck coefficient,  mean p-type doping, which can be realized by shifting the Fermi level into valence bands.
The biaxial strain dependence of S,  $\mathrm{\sigma/\tau}$ and $\mathrm{S^2\sigma/\tau}$ using GGA+SOC at room temperature  are plotted in \autoref{t3}. The energy band structures  of $\mathrm{PtSe_2}$ is  sensitively dependent on strain, which leads to
complex strain  dependence  of  transport coefficients. In n-type doping, compressive strain induces larger Seebeck coefficient (absolute values), while tensile strain has little effects on Seebeck coefficient. In p-type doping, the Seebeck coefficient firstly increases, and then decreases with increasing $a$. As $a$ increases, the  $\mathrm{\sigma/\tau}$ firstly increases, and then decreases for n-type, while the opposite trend is observed for p-type. Considering  the comprehensive  strain effects on S and  $\mathrm{\sigma/\tau}$, compressive strain can significantly enhance the n-type power factor, while tensile strain can greatly  improve the p-type power factor. The similar strain effects on power factor also can be found in monolayer $\mathrm{MoS_2}$\cite{gsd1}.

Strain-enhanced power factor can be explained by strain-driven  accidental degeneracies, namely bands converge.
In considered n-type doping range, the largest S  can be attained with $a$=3.63  $\mathrm{{\AA}}$ among considered $a$ due to the near degeneracy  between conduction band extremum along   K-$\Gamma$  and one  along  $\Gamma$-M, leading to largest power factor. For p-type,  S  reaches the peak with $a$=3.78  $\mathrm{{\AA}}$, because the energy levels of one point along  K-$\Gamma$ line and $\Gamma$ point are more adjacent. However, the lowest  $\mathrm{\sigma/\tau}$ is attained  with $a$=3.78  $\mathrm{{\AA}}$ due to more localized first valence band.  The largest p-type power factor can be attained with $a$=3.88  $\mathrm{{\AA}}$ due to the largest $\mathrm{\sigma/\tau}$ and  relatively large S.
When  strain is applied, the CBM or VBM transition is induced, and the corresponding critical $a$ can produce the larger  S in the considered $a$ and doping range, which is   beneficial to power factor.
Calculated results show  that strain-enhanced n-type  power factor by using compressive strain is larger than p-type  one by applied tensile strain.
\begin{figure*}
  \includegraphics[width=14cm]{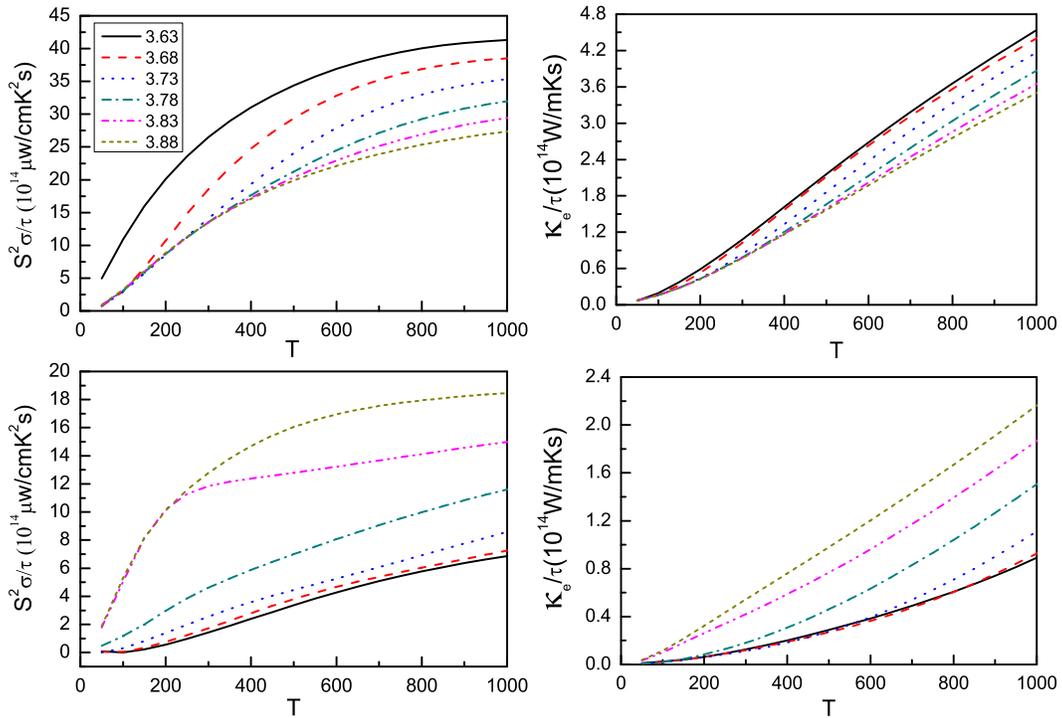}
  \caption{(Color online) For both n-type (Top) and p-type (Bottom), the  power factor with respect to scattering time $\mathrm{S^2\sigma/\tau}$ and  electronic thermal conductivity  with respect to scattering time  $\mathrm{\kappa_e/\tau}$ as a function of temperature  with $a$ changing from 3.63 $\mathrm{{\AA}}$ to 3.88 $\mathrm{{\AA}}$  using GGA+SOC, and the doping concentration is $\mathrm{2.012\times10^{13}cm^{-2}}$ (about 0.025 electrons or holes per unit cell).}\label{t5}
\end{figure*}
\begin{figure*}
    \includegraphics[width=16.5cm]{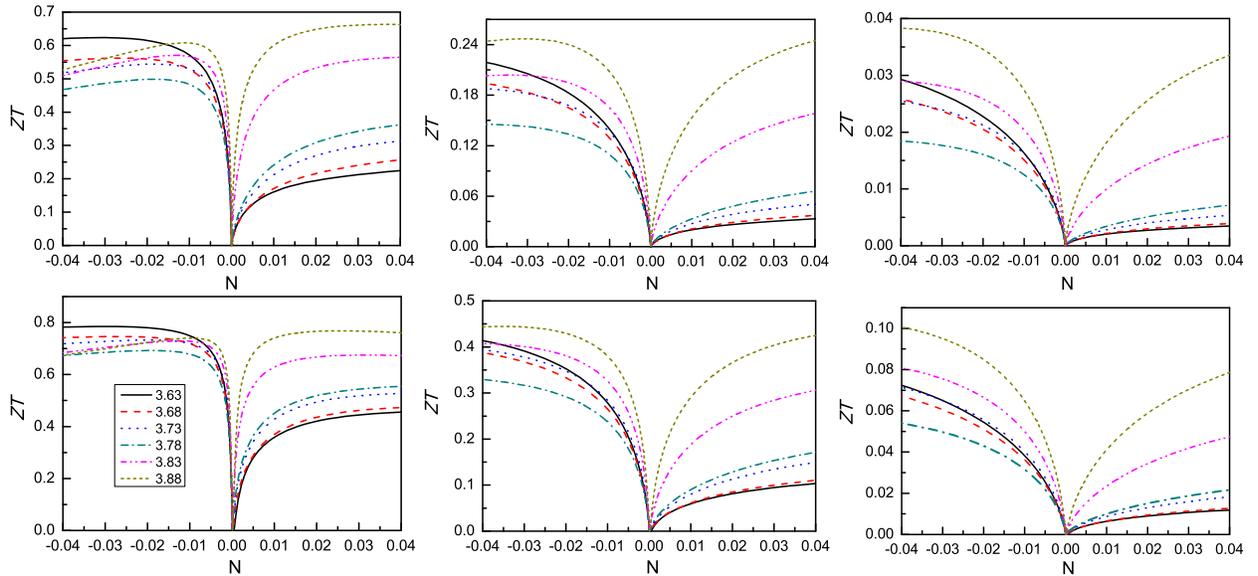}
  \caption{(Color online) At 600 K (Top) and 900 K (Bottom), the $ZT$  as a function of doping level  with $a$ changing from 3.63 $\mathrm{{\AA}}$ to 3.88 $\mathrm{{\AA}}$, and the scattering time $\mathrm{\tau}$  is 1 $\times$ $10^{-13}$ s (Left), 1 $\times$ $10^{-14}$ s (Middle) and 1 $\times$ $10^{-15}$ s (Right).}\label{t6}
\end{figure*}

Another key parameter of thermoelectric materials is thermal conductivity, including electronic and lattice thermal conductivities.   The strain dependence of  electronic thermal conductivity   with respect to scattering time  $\mathrm{\kappa_e/\tau}$ (300 K) as a function of doping level and lattice  thermal conductivity as a function of temperature   are plotted in \autoref{t4}.  The $\mathrm{S^2\sigma/\tau}$ has similar strain dependence with power factor.
We assume that the lattice thermal conductivity is independent of  doping level, and it  typically goes as 1/T at high temperature. It is found that tensile strain can induce lower  lattice thermal conductivity, which  is very  beneficial to the efficiency of thermoelectric conversion.
The room temperature lattice  thermal conductivity (6.88 $\mathrm{W m^{-1} K^{-1}}$) with  $a$=3.88  $\mathrm{{\AA}}$   is about 60\%  smaller than that (16.97 $\mathrm{W m^{-1} K^{-1}}$) with optimized lattice constant  $a$=3.73  $\mathrm{{\AA}}$.

Finally, the $\mathrm{S^2\sigma/\tau}$ and $\mathrm{\kappa_e/\tau}$ as a function of temperature  with the doping concentration of  $\mathrm{2.012\times10^{13}cm^{-2}}$ for both n- and p-type  are shown in \autoref{t5}.
In the considered  temperature range,  the strain dependence of both $\mathrm{S^2\sigma/\tau}$ and $\mathrm{\kappa_e/\tau}$
is consistent with one at 300 K. The n-type  power factor with $a$=3.63  $\mathrm{{\AA}}$  and p-type one with $a$=3.88  $\mathrm{{\AA}}$  are the largest among the considered $a$. To attain figure of merit $ZT$,  the scattering time $\mathrm{\tau}$ is unknown. Calculating scattering time $\mathrm{\tau}$ is challenging from the first-principle calculations due to the complexity of various carrier scattering mechanisms. To attain possible $ZT$ values, the scattering time $\mathrm{\tau}$ is assumed to be
1 $\times$ $10^{-13}$ s, 1 $\times$ $10^{-14}$ s  and 1 $\times$ $10^{-15}$ s. In Ref.\cite{t5}, the scattering time of monolayer $\mathrm{MoS_2}$ is fitted as 2.29 $\sim$ 5.17 $\times$ $10^{-14}$ s to calculate $ZT$. For $\mathrm{WSe_2}$, the scattering time
is found to be 1.6 $\times$ $10^{-13}$ or  1.4 $\times$ $10^{-15}$ s\cite{t6}. Therefore, our assumed scattering time should be
 reasonable.  At 600 K and 900 K, the $ZT$  as a function of doping level  with $a$ changing from 3.63 $\mathrm{{\AA}}$ to 3.88 $\mathrm{{\AA}}$ are plotted in \autoref{t6}. The similar strain dependence between 600 K and 900 K is observed except for  relative sizes of $ZT$. It is found that the $ZT$ decreases with decreasing $\mathrm{\tau}$, which is because the larger $\mathrm{\tau}$ produces larger power factor.
In p-type doping, tensile strain can observably improve the $ZT$ for all three $\mathrm{\tau}$. For n-type, tensile strain can also enhance  the $ZT$ with  $\mathrm{\tau}$=1 $\times$ $10^{-14}$ s  and  $\mathrm{\tau}$=1 $\times$ $10^{-15}$ s. However, compressive strain can slightly improve $ZT$ with  $\mathrm{\tau}$=1 $\times$ $10^{-13}$ s. The peak $ZT$ is about 0.65, 0.25 and 0.04 with decreasing  $\mathrm{\tau}$ at 600 K, and  0.80, 0.45 and 0.1 at 900 K.
 Calculated results show that tensile strain may be a effective method to attain higher $ZT$, which can achieve higher  thermoelectric conversion efficiency.

\begin{figure}
  \includegraphics[width=8cm]{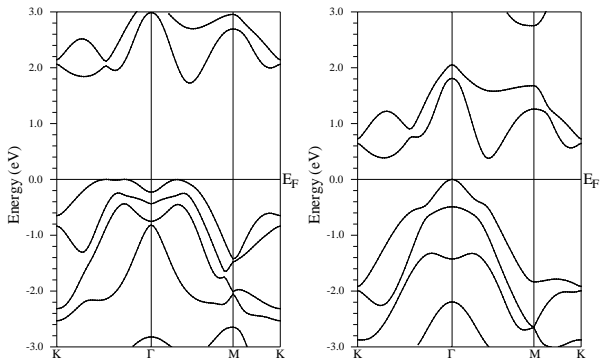}
\caption{The energy band structures  of monolayer $\mathrm{PtS_2}$ (Left) and  $\mathrm{PtTe_2}$ (Right) with unstrained lattice constants  using GGA+SOC. }\label{t7}
\end{figure}

\section{Discussions and Conclusion}
The semiconducting transition-metal dichalcogenide monolayers may be potential thermoelectric materials. However,
huge works focused on thermoelectric properties of monolayer $\mathrm{MoS_2}$. In Ref.\cite{gsd2}, we systematically investigated electronic transport properties of semiconducting transition-metal dichalcogenide monolayers $\mathrm{MX_2}$ (M=Zr, Hf, Mo, W and Pt; X=S, Se and Te), including SOC.  Among all cation groups, Pt cation group exhibits best power factor due to the highest Seebeck coefficient, assuming  scattering time to be fixed. Here, recent synthetic $\mathrm{PtSe_2}$ is investigated, whose  room temperature lattice thermal conductivity  (16.97 $\mathrm{W m^{-1} K^{-1}}$) is lower  than  one (26.2 $\mathrm{W m^{-1} K^{-1}}$\cite{t3}) of monolayer $\mathrm{MoS_2}$ with the similar calculation method. So, monolayer $\mathrm{PtSe_2}$ may possess better thermoelectric properties than monolayer $\mathrm{MoS_2}$. The high lattice thermal conductivity is a major disadvantage to obtain higher $ZT$. However, the lattice thermal conductivity can be reduced by phonon engineering, such as isotope doping\cite{lc1}, nanoporous
structure\cite{lc2}  or strain\cite{lc3}. The typical example is graphene, whose lattice thermal conductivity  can be reduced largely by phonon engineering, producing  a very high  $ZT$ of 3\cite{lc4}.  The pressure-reduced lattice thermal conductivity also can be found in $\mathrm{Mg_2Sn}$\cite{lc3}. Here, tensile strain can induce remarkably   reduced lattice thermal conductivity, from 16.97 $\mathrm{W m^{-1} K^{-1}}$ to 6.88 $\mathrm{W m^{-1} K^{-1}}$ at 300 K with $a$ changing from 3.73  $\mathrm{{\AA}}$ to  $a$=3.88  $\mathrm{{\AA}}$ , and the corresponding $\varepsilon$  is about -4.02\%, which should be easily achieved in experiment by piezoelectric stretching and exploiting the thermal expansion mismatch\cite{bs1,bs2}.

The electronic structures of semiconducting transition-metal dichalcogenide monolayers is quite sensitive to strain, which provides a strategy  to tune their  thermoelectric properties by band engineering.
Strain or pressure is a conventional  way to induce novel phenomenon, such as pressure-induced high-Tc superconductivity\cite{ht1,ht2} and strain-induced topological insulator\cite{tps1}. The  symmetry-driven degeneracy, low-dimensional electronic structures and accidental degeneracies are three usual mechanisms to induce high Seebeck coefficient suitable for high power factor.
Here, strain-induced accidental degeneracies, namely band convergence, can be used to explain strain-enhanced Seebeck coefficient.
For optimized lattice constants $a$=3.73  $\mathrm{{\AA}}$, monolayer $\mathrm{PtSe_2}$  has some  valence band extrema (VBE) and conduction band extrema (CBE) around the Fermi level, which provides a platform to achieve band convergence by strain.
When compressive strain gradually increases, the CBE along K-$\Gamma$ and $\Gamma$-M  approach each other, and the energy difference changes from 0.140 eV  to 0.009 eV with $a$ being 3.73  $\mathrm{{\AA}}$ to 3.63  $\mathrm{{\AA}}$.
The conduction band convergence produces large n-type Seebeck coefficient, giving rise to high n-type power factor. When tensile  strain gradually increases,  the VBE along K-$\Gamma$ and  VBM   are more close,  and the energy difference varies from 0.184 eV  to 0.062 eV with $a$ changing from 3.73  $\mathrm{{\AA}}$ to 3.78  $\mathrm{{\AA}}$. The valence band convergence induces large  p-type Seebeck coefficient. As the $a$ continues to increase, the extrema at $\Gamma$ point disappears, and another extrema along $\Gamma$-M appears, which induces significantly enhanced p-type electrical conductivity. The largest p-type power factor achieves at $a$=3.88  $\mathrm{{\AA}}$. Calculated results show that the large  Seebeck coefficient  can be induced by both compressive and tensile strain at the critical strain of CBM or VBM transition.
Similar pressure or strain  induced band convergence, leading to large Seebeck coefficient, also can be found in $\mathrm{Mg_2Sn}$\cite{lc3} at the critical pressure of energy band gap or monolayer $\mathrm{MoS_2}$ at the critical strain of direct-indirect gap transition\cite{gsd1}. The n-type doping related results  indicates that a large
power factor will not certainly produce  a high $ZT$, while a moderate power factor  combined with a suitable  thermal conductivity may eventually  lead to  a high $ZT$.

\begin{figure}
  \includegraphics[width=7cm]{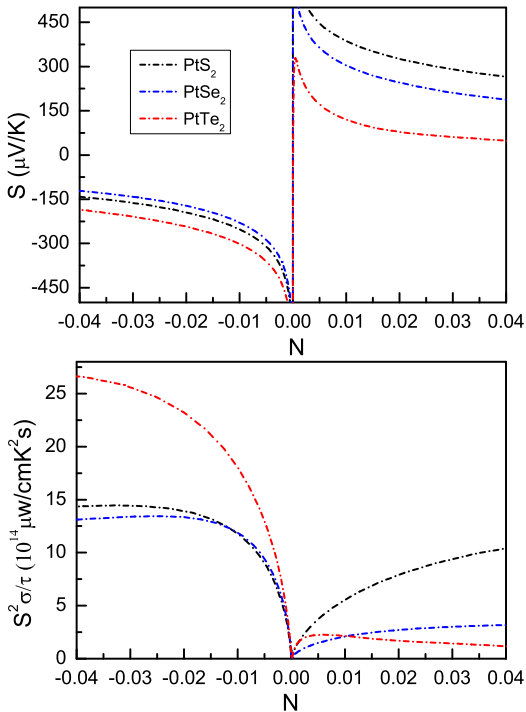}
\caption{(Color online) At 300 K, transport coefficients of  $\mathrm{PtX_2}$ (X=S, Se and Te)) with unstrained lattice constants, including  Seebeck coefficient S (Top) and  power factor with respect to scattering time $\mathrm{S^2\sigma/\tau}$ (Bottom),  as a function of doping level (N) using GGA+SOC. }\label{t8}
\end{figure}

In fact, band convergence can be observed in  unstrained $\mathrm{PtS_2}$ and  $\mathrm{PtTe_2}$, and the energy band structures  of monolayer $\mathrm{PtS_2}$  and  $\mathrm{PtTe_2}$  with unstrained lattice constants 3.57  $\mathrm{{\AA}}$ and 4.02  $\mathrm{{\AA}}$ using GGA+SOC are plotted in \autoref{t7}. For $\mathrm{PtS_2}$, valence band convergence can be observed, and the similar results for $\mathrm{PtSe_2}$ can be achieved by tensile strain.  For $\mathrm{PtTe_2}$, conduction  band convergence can be seen, which can be attained for $\mathrm{PtSe_2}$ by compressive strain. Band convergence is favorable for Seebeck coefficient, leading to high power factor. To clearly illustrate these results,  the room temperature transport coefficients of  $\mathrm{PtX_2}$ (X=S, Se and Te) with unstrained lattice constants, including  Seebeck coefficient S  and  power factor with respect to scattering time $\mathrm{S^2\sigma/\tau}$,  as a function of doping level using GGA+SOC are shown in \autoref{t8}.
It is very  clear to see that $\mathrm{PtS_2}$ ($\mathrm{PtTe_2}$) has the largest p-type (n-type) Seebeck coefficient, which is
 consistent with corresponding band convergence. It is resultant  that $\mathrm{PtS_2}$ ($\mathrm{PtTe_2}$) has the highest p-type (n-type) power factor.

 In summary,  we  systematically study strain dependence of  thermoelectric properties  of monolayer $\mathrm{PtSe_2}$, including both  electron and phonon transport,  using GGA+SOC, based mainly on the reliable first-principle calculations.
 It is found that both compressive and tensile strain can induce improved Seebeck coefficient
at the critical strain of CBM or VBM transition, which is favorable for power factor. Calculated results also show that tensile strain can lead to significantly reduced lattice thermal conductivity, which is beneficial to $ZT$.
By using hypothetical  scattering time $\mathrm{\tau}$, $ZT$ can be obtained, which shows that strain indeed can induce enhanced  efficiency of thermoelectric conversion due to improved $ZT$ value. So, strain is a very effective method to achieve
enhanced thermoelectric properties for monolayer $\mathrm{PtSe_2}$, which provides great opportunities for efficient thermoelectricity. The strategy of strain-tuned thermoelectric properties also can be used in other semiconducting transition-metal dichalcogenide monolayers, like  $\mathrm{PtS_2}$ and  $\mathrm{PtTe_2}$ with high power factor.

\begin{acknowledgments}
This work is supported by the National Natural Science Foundation of China (Grant No. 11404391). We are grateful to the Advanced Analysis and Computation Center of CUMT for the award of CPU hours to accomplish this work.
\end{acknowledgments}

\end{document}